\documentstyle[twoside,fleqn,espcrc2,epsfig]{article}
\def \3{\ss }
\newcommand{\half}{\mbox{\small $\frac{1}{2}$}}
\newcommand{\third}{\mbox{\small $\frac{1}{3}$}}
\newcommand{\cO}{{\cal O}}
\newcommand{\Dd}[1]{\mbox{
  \parbox[b]{0cm}{$D$}\raisebox{1.7ex}{$\leftrightarrow$}$_{\!#1}$}}

\begin{document}
 
\title{\vspace{-3.65cm}
       {\normalsize DESY 96--151}    \\[-0.2cm]
       {\normalsize HLRZ 96--55}     \\[-0.2cm]
       {\normalsize HUB--EP--96/38}  \\[-0.2cm]
       {\normalsize August 1996}   \\
       \vspace{1.55cm}
        Lattice computation of structure functions\thanks{Plenary talk
 at Lattice 96, St.\ Louis, presented by M. G\"ockeler.}}

\author{M. G\"ockeler\address{Institut f\"ur Theoretische Physik,
     J.W. Goethe-Universit\"at, D-60054 Frankfurt(Main), Germany}$^,$%
\address{Institut f\"ur Theoretische Physik E,
                         RWTH Aachen, D-52056 Aachen, Germany}$^,$%
\address{H{\"o}chstleistungsrechenzentrum HLRZ, 
 c/o Forschungszentrum J{\"u}lich, D-52425 J{\"u}lich, Germany},
R. Horsley\address{Institut f\"ur Physik, Humboldt-Universit\"at zu Berlin, 
      Invalidenstr.\ 110, D-10115 Berlin, Germany}, 
E.-M. Ilgenfritz$^{\rm d}$,
H. Perlt\address{Institut f\"ur Theoretische Physik, Universit\"at 
      Leipzig, Augustusplatz 10-11, D-04109 Leipzig, Germany},
P. Rakow$^{\rm c}$,
G. Schierholz\address{Deutsches Elektronen-Synchrotron DESY,
      Notkestr.\ 85, D-22603 Hamburg, Germany}$^{\rm ,c}$ and
A. Schiller$^{\rm e}$}

\begin{abstract}
Recent lattice calculations of hadron structure functions are described.
\end{abstract}
 
\maketitle
 
\section{INTRODUCTION}
Since the late 1960's, deep-inelastic scattering of unpolarised
leptons and nucleons has been studied experimentally. The inclusive
cross section for electroproduction can be described in terms of the 
structure functions $F_1$ and $F_2$ related to the densities of
quarks and gluons in the nucleon (see, e.g., Ref.\ \cite{soper}).
There is also some more indirect information on pion
structure functions.
Deep-inelastic scattering of longitudinally polarised leptons and
nucleons 
allows one to measure two more structure functions, $g_1$ and $g_2$. 
 
The observed scaling violations 
could be understood in the framework of {\em perturbative}
QCD 
The computation of the structure functions themselves
requires a {\em nonperturbative} method, e.g.\ lattice gauge theory 
and Monte Carlo simulations 
\cite{martinelli}. In this talk I shall describe recent 
attempts to calculate
hadron structure functions on the lattice. An alternative method
using the Hamiltonian version of lattice gauge theory is discussed in
Ref.\ \cite{scheu}.

Instead of calculating the structure functions directly, we compute
hadronic matrix elements of certain operators which, through the 
operator product expansion, are related to moments 
of the structure functions. In the deep-inelastic limit, 
the leading contribution is given by
\begin{equation}
2 \int_0^1 dx \, x^{n-1} F_1(x,Q^2) = \sum_f c^{(f)}_{1,n}   
    v^{(f)}_n (\mu) \,,
\end{equation}
\begin{equation}
\int_0^1 dx \, x^{n-2} F_2(x,Q^2) = \sum_f c^{(f)}_{2,n}   
    v^{(f)}_n (\mu) \,.
\end{equation}
Here $n=2,4,6,\ldots$; $f$ labels the contributing operators, and
$\mu$ is the renormalisation scale needed to define the (reduced)
matrix elements $v^{(f)}_n$. The Wilson coefficients $c^{(f)}_{i,n}$
are calculated in perturbation theory and depend on $\mu^2 / Q^2$ as
well as on the coupling constant $g(\mu)$. 

The operators we have to deal with are the ``2-quark operators'' (for
$f=q=u,d$)  
\begin{equation}
  \cO ^{(q)}_{\mu_1 \cdots \mu_n} =   
       \left(\frac{\mbox{i}}{2}\right)^{n-1} \bar{q}\gamma_{\mu_1} 
          \Dd{\mu_2} \cdots \Dd{\mu_n} q 
\end{equation}
($D$ = covariant derivative)
and (for $f=g$) purely gluonic operators,
which contribute only in the flavour-singlet sector. 
For the nucleon the matrix elements
$v^{(f)}_n$ are then defined by 
\begin{equation} \begin{array}{l} \displaystyle
  \half \sum_s \langle p,s | \cO ^{(f)}_{\{\mu_1 \cdots \mu_n\}} 
     - \mbox{traces} | p,s \rangle 
\\ \displaystyle
\quad {} = 2 v^{(f)}_n (p_{\mu_1} \cdots p_{\mu_n}
     - \mbox{traces} ) \,,
\end{array}
\end{equation}
where $|p,s\rangle$ denotes a nucleon state with momentum $p$ 
and spin vector $s$ ($s^2 = - m^2$). 
Symmetrisation of all indices (indicated by $\{ \cdots \}$) and
subtraction of traces are needed to obtain operators transforming
irreducibly under the Lorentz group,
i.e.\ operators of definite twist = 2. 
Analogous results exist for pion and rho as well.

Within the parton model the $v^{(f)}_n$ are
interpreted as average values of powers of the fraction of the hadron
momentum carried by the parton (quark of flavour $q$ or gluon):
\begin{equation}
  v^{(f)}_n = \langle x^{n-1} \rangle ^{(f)} \,. 
\end{equation}

For the polarised nucleon structure functions one obtains in the
chiral limit:
\begin{equation}
2 \int_0^1 dx \, x^n g_1(x,Q^2) = \half \sum_f e^{(f)}_{1,n}   
    a^{(f)}_n (\mu) 
\end{equation}
for $n=0,2,4,\ldots$ and
\begin{equation} \begin{array}{l} \displaystyle
2 \int_0^1 dx \, x^n g_2(x,Q^2) 
\\ \displaystyle
 {} = \half \frac{n}{n+1}
    \sum_f \left[ e^{(f)}_{2,n} d^{(f)}_n (\mu) - 
e^{(f)}_{1,n} a^{(f)}_n (\mu) \right] 
\end{array}
\end{equation}
for $n=2,4,\ldots$
with Wilson coefficients $e^{(f)}_{i,n}$ and reduced matrix elements
defined by \cite{jaffe}
\renewcommand{\arraystretch}{1.5}
\begin{equation} \begin{array}{l} \displaystyle
 \langle p,s | \cO ^{5(f)}_{\{\sigma \mu_1 \cdots \mu_n\}} 
     - \mbox{traces} | p,s \rangle 
\\ \displaystyle
 {} = \frac{a^{(f)}_n}{n+1} \left(s_\sigma p_{\mu_1} \cdots p_{\mu_n}
     + \cdots - \mbox{traces} \right) \,,
\end{array}
\end{equation}
\begin{equation} \begin{array}{l} \displaystyle
 \langle p,s | \cO ^{5(f)}_{[\sigma \{\mu_1] \mu_2 \cdots \mu_n\}} 
     - \mbox{traces} | p,s \rangle 
\\ \displaystyle
 {} = \frac{d^{(f)}_n}{n+1} 
    ( (s_\sigma p_{\mu_1} -s_{\mu_1} p_\sigma ) p_{\mu_2} 
     \cdots p_{\mu_n} + \cdots 
\\ \displaystyle
     \hphantom{=} {} - \mbox{traces} ) \,. 
\end{array}
\end{equation}
Here we need the operators (for $f=q$)
\begin{equation}
  \cO ^{5(q)}_{\sigma \mu_1 \cdots \mu_n} =   
       \left(\frac{\mbox{i}}{2}\right)^n \bar{q}\gamma_\sigma \gamma_5
          \Dd{\mu_1} \cdots \Dd{\mu_n} q  \,.
\end{equation}
Note that $d^{(f)}_n$ corresponds to twist 3.
Within the parton model, 
$\half a^{(q)}_0 \equiv \Delta q$ is interpreted as
the fraction of the nucleon spin
carried by the quarks of flavour $q$.

Now the challenge for lattice QCD is the calculation of the reduced
matrix elements $v_n$, $a_n$, $d_n$.
So we have to compute forward hadron matrix
elements of (relatively complicated) composite operators.

\section{LATTICE CALCULATION}

These matrix elements are calculated from three-point correlation 
functions. With suitable
operators $B$, $\bar{B}$ for the particle to be studied, e.g.\ the
nucleon, we can write schematically for
$t > \tau > 0$ 
\begin{equation} \begin{array}{l} \displaystyle
  \langle B(t) {\cal O} (\tau) \bar{B}(0) \rangle 
\\ \displaystyle
  {} = \langle 0 |  B \mbox{e}^{- H (t-\tau)} {\cal O}
       \mbox{e}^{- H \tau} \bar{B} | 0 \rangle
\\ \displaystyle
  {} = \langle 0 |  B | N \rangle \mbox{e}^{- E_N t }
       \langle N | \bar{B} | 0 \rangle
       \langle N | {\cal O}| N \rangle + \cdots
\end{array}
\end{equation}
on a lattice with time extent $T \to \infty$. Correspondingly we have
for the two-point function 
\begin{equation} \begin{array}{l} \displaystyle
  \langle B(t) \bar{B}(0) \rangle =
  \langle 0 |  B \mbox{e}^{- H t}  \bar{B} | 0 \rangle
\\ \displaystyle
  {} \stackrel{t \to \infty}{=} 
       \langle 0 |  B | N \rangle \mbox{e}^{- E_N t}
       \langle N | \bar{B} | 0 \rangle + \cdots
\end{array}
\end{equation}
So we can determine the desired  matrix elements from ratios of
the form
\begin{equation} \label{ratio}
 R \equiv  
 \frac{ \langle B(t) {\cal O} (\tau) \bar{B}(0) \rangle}
      { \langle B(t) \bar{B}(0) \rangle}
  = \langle N | {\cal O}| N \rangle + \cdots \,,
\end{equation}
which should be independent of $\tau$ for $ 0 \ll \tau \ll t $.
 
The bare lattice operators have of course to be renormalised and may
mix with other operators in the process of renormalisation. 
In the euclidean continuum we should study operators like 
\begin{equation}
  \cO ^{(q)}_{\mu_1 \cdots \mu_n} =   
        \bar{q}\gamma_{\mu_1} \Dd{\mu_2} \cdots \Dd{\mu_n} q \,, 
\end{equation}
\begin{equation}
  \cO ^{5(q)}_{\sigma \mu_1 \cdots \mu_n} =   
        \bar{q}\gamma_\sigma \gamma_5 \Dd{\mu_1} \cdots \Dd{\mu_n} q 
\end{equation}
or rather O(4) irreducible multiplets with definite C-parity. In
particular, we obtain twist-2 operators by symmetrising the indices
and subtracting the traces. In the flavour-nonsinglet case they do not
mix and are hence multiplicatively renormalisable. 

Working with Wilson fermions (as we do) it is straightforward to write
down lattice versions of the above operators. One simply replaces the
continuum covariant derivative by its lattice analogue. However, O(4)
being restricted to its finite subgroup H(4) (the hypercubic group) on the
lattice, the constraints imposed by space-time symmetry are less
stringent than in the continuum. 
In particular, a multiplet of operators which is irreducible
with respect to O(4) will in general decompose into {\em several}
irreducible H(4) multiplets and the possibilities for mixing increase
\cite{grouptheory,pertz}.

\section{THE SIMULATIONS}

Our simulations were performed on Quadrics parallel computers. 
We have worked on lattices of size $L^3 \times T = 16^3 \times 32$ and 
$24^3 \times 32$ at $\beta = 6.0$ in the quenched approximation. 
For more technical details see Refs.\ \cite{letter,biele}.
Most of our data were
obtained with the Wilson action for the gauge fields and the quarks. 
First results from simulations with an improved
fermionic action will however be discussed
below (for more details see Ref.\ \cite{stephenson}). 

The values of $\kappa$, $L$ and the (approximate) number of
configurations used for the calculation of the
nucleon matrix elements are the following :
\renewcommand{\arraystretch}{1.0}
\begin{center}
\begin{tabular}{r@{.}lcc}
 \multicolumn{2}{c}{$\kappa$}  &  $L$        & configs.   \\ 
             0    &   1515     &  16         & 400        \\
             0    &   153      &  16         & 600        \\
             0    &   155      &  16         & 900        \\
             0    &   155      &  24         & 100        \\
             0    &   1558     &  24         & 100        \\
             0    &   1563     &  24         & 100     
\end{tabular}
\end{center}
Comparison of the $L=16$ and $L=24$ lattices at $\kappa = 0.155$ will
give us some information on finite size effects. 

On the larger lattice we can
use lighter quark masses without running 
into problems from severe finite size effects. This makes the
extrapolation to the chiral limit more reliable, but 
the time needed to invert the fermion matrix increases considerably. 
So it becomes important that we choose our algorithm carefully. 
Therefore we have made a comparison between two popular choices, minimal
residue with overrelaxation (MR) and  a stabilised variant of the 
biconjugate gradient algorithm (BiCGstab).  

Our main findings (for $\beta=6.0$, Wilson action quarks) are that 
if the quarks are not very light the MR algorithm, with an overrelaxation
parameter $\omega = 1.1$ was the most efficient algorithm, saving about
15 \% in CPU time compared with BiCGstab. However as we approach 
$\kappa_c$ BiCGstab becomes the preferred algorithm. 
The $\kappa$ where the two inversion times
cross over depends strongly on the lattice size. On a $12^3 \times 16$ 
lattice the crossover is near $\kappa \approx 0.154$ while on the 
$24^3 \times 32$ lattice BiCGstab does not become the better algorithm 
until $\kappa \approx 0.1555$. For our calculations we have therefore
used the MR algorithm for the lower $\kappa$ values, the BiCGstab for the
highest.  

In order to suppress the unwanted excited states as much as possible
it is important to choose the hadron operators judiciously. We do this
by applying Jacobi smearing to the standard local operators. Both
source and sink are smeared, since one needs a good projection on the
ground state on both sides of the inserted operator in the three-point
function. In the case of the proton we additionally use the
``nonrelativistic projection'' \cite{letter}. 
In this way we obtain a sufficiently long interval in $t$, 
where the two-point function is dominated by the ground state. 

Our final choice of the operators whose matrix elements are calculated
is motivated by the wish to avoid mixing (as far as possible) as well as
momenta with more than one nonzero component. Taking the nucleon
polarisation (where needed) in 2-direction and choosing the momenta
$\vec{p} = (0,0,0),(2 \pi /L,0,0)$ we studied the following operators  
and reduced matrix elements.
\begin{center} 
  \begin{tabular}{cc}   \hline
    $v_{2,a}$  &  $ \cO ^{(q)}_{\{14\}} $  \\
    $v_{2,b}$  &  $ \cO ^{(q)}_{\{44\}} - \third ( \cO ^{(q)}_{\{11\}}
                    + \cO ^{(q)}_{\{22\}} + \cO ^{(q)}_{\{33\}} ) $     \\    
    $v_3$      &  $ \cO ^{(q)}_{\{114\}} - \half ( \cO ^{(q)}_{\{224\}}
                    + \cO ^{(q)}_{\{334\}} ) $     \\    
    $v_4$      &  $ \cO ^{(q)}_{\{1144\}} + \cO ^{(q)}_{\{2233\}}
                    - \cO ^{(q)}_{\{1133\}} - \cO ^{(q)}_{\{2244\}}  $  \\    
    $a_0$      &  $ \cO ^{5(q)}_2 $  \\
    $a_2$      &  $ \cO ^{5(q)}_{\{214\}} $   \\
    $d_2$      &  $ \cO ^{5(q)}_{[2\{1]4\}} $ \\ \hline
  \end{tabular}
\end{center}
The two operators for $v_2$ belong to different representations of
H(4). Therefore their comparison 
gives an indication of the size of lattice artifacts.  

Recently, first results have been obtained for nucleon matrix elements
of operators of the type
\begin{equation}
 \bar{q}\sigma_{\sigma \mu_1} \gamma_5 \Dd{\mu_2} \cdots \Dd{\mu_n} q \,,
\end{equation}
which are related to the so-called transversity distribution $h_1$
\cite{h1}. Purely gluonic operators have also been
studied although they fluctuate much more than the 2-quark operators
\cite{horsley}. 

As explained above, we extract the hadronic matrix elements of our
operators from the ratios $R$ (see (\ref{ratio})).
In the case of the 2-quark operators we fix $t$, which
allows us to compute the quark-line connected part of the 
three-point function for all values of $\tau$ and arbitrary operators 
$\cO$ from quark propagators on two sources. The disconnected
insertions, which would contribute in the flavour-singlet sector only,
are however omitted. We choose $t$ as large as possible in order to
have enough space for a ``plateau'' between 0 and $t$ where $R$ is
independent of $\tau$. For the nucleon we took $t/a=13$ whereas in the
case of the pion we adopted the symmetrical choice $t/a=16$
on our $16^3 \times 32$ lattices.  

\begin{figure}[t]   
\vspace*{-0.8cm}
\epsfig{file=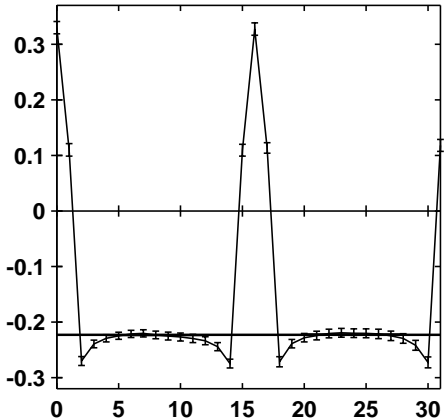,width=7.5cm} 
\vspace*{-1.7cm}
 \caption[dummy]{$R$ for the pion matrix element $v_{2,b}$ at
                 $\kappa = 0.153$ as a function of $\tau /a$.}
\label{fig:rpion}
\end{figure}

Indeed we observe reasonable plateaus.
Examples for the nucleon are shown in Ref.\ \cite{letter}, an example
for the pion is plotted in Fig.\ref{fig:rpion}.
From the values of $R$ on the plateaus we can then
calculate the reduced matrix elements 
taking into account kinematical factors, renormalisation constants
etc.\ \cite{letter}.
\begin{figure}[t]   
\vspace*{-0.6cm}
\hspace*{-1cm}
\epsfig{file=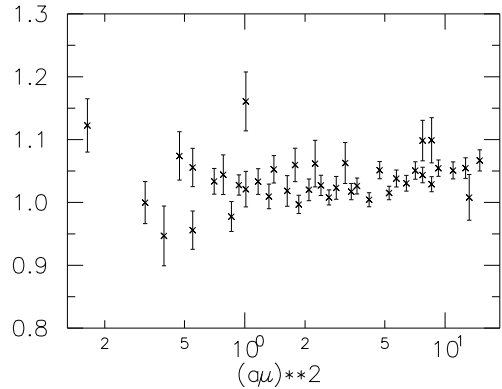,width=7.5cm,angle=90} 
\vspace*{-2.7cm}
 \caption[dummy]{$Z \times$Wilson coefficient for the  matrix 
            element $v_{2,a}$ at $\kappa = 0.153$.}
\label{fig:zplot}
\end{figure}
\setlength{\textfloatsep}{0.7cm}

\section{RENORMALISATION}

Using 1-loop lattice perturbation theory we have calculated the
matrix elements of all operators
$\cO^{(q)}_{\mu_1 \cdots \mu_n}$ ($n \leq 4$) and 
$\cO^{5(q)}_{\sigma \mu_1 \cdots \mu_n}$ ($n \leq 2$) between quark
states in the quenched approximation \cite{pertpaper}. 
From these one can immediately compute the
renormalisation constants and mixing coefficients for any linear
combination that one wants to study. 
Increasing the accuracy of the required numerical integrations
we obtained up to 8 significant digits.
In the cases considered
previously \cite{pertz}, our results agree with the older ones.
In the following, we shall use the perturbative $Z$'s at the scale 
$\mu^2 = a^{-2} \approx 5 \mbox{GeV}^2$.

However, a nonperturbative determination is
possible \cite{nonpertz} and should eventually be preferred.
In the end, the $\mu$ dependence of $Z$ must be cancelled by the $\mu$
dependence of the Wilson coefficient leading to renormalisation
prescription independent results for the structure functions. In order
to show to which extent this matching can be achieved 
we plot in Fig.\ref{fig:zplot} the
product of the nonperturbatively calculated $Z$ for $v_{2,a}$ with the
corresponding renormalisation group improved nonsinglet
Wilson coefficient versus
$a^2 \mu^2$. Indeed we obtain a reasonably flat region around 
$a^2 \mu^2 \approx 2$, though with a rather low value of 
$\Lambda_{\rm QCD} \approx 100 {\rm MeV}$.

\section{RESULTS FOR UNPOLARISED NUCLEON STRUCTURE FUNCTIONS}

\begin{figure}[t]   
\vspace*{-1.4cm}
\hspace*{-1cm}
\epsfig{file=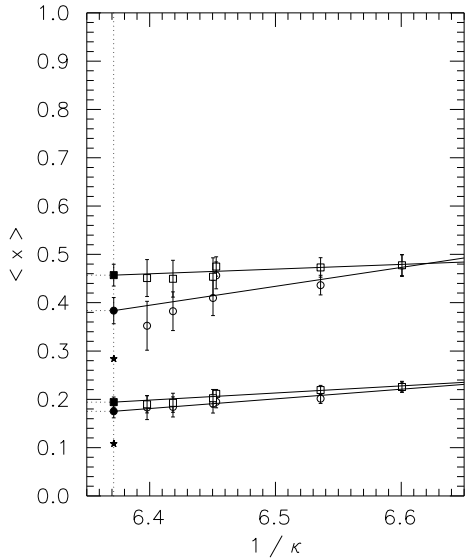,width=9cm}
 \vspace*{-4.2cm}
 \caption[dummy]{$\langle x \rangle$ for the proton (MOM scheme). 
       The circles (boxes) correspond to $v_{2,a}$ ($v_{2,b}$).
       The upper (lower) band of data represents
       the results for the up (down)-quark distribution.}
\label{fig:proton}
\end{figure}

Plotting our results for $v_2$, $v_3$, $v_4$ versus $1/ \kappa$ we can
extrapolate (linearly) to the chiral limit. This is shown in 
Fig.\ref{fig:proton} for
$v_2 \equiv \langle x \rangle$ in the proton. (Similar pictures for
$v_3$ and $v_4$ can be found in Ref.\ \cite{letter}.)
We see that the values obtained at smaller quark masses on
the larger lattice ($24^3 \times 32$) are consistent with those coming
from the $16^3 \times 32$ lattice. In particular, at $\kappa = 0.155$ 
the numbers from both volumes agree within the errors. Furthermore we
observe approximate consistency among $v_{2,a}$ and $v_{2,b}$ 
indicating the absence of large lattice artifacts
(at least in this case). 

At this point it is impossible to resist the temptation to compare our
results with experimental numbers. However, due to our use of the
quenched approximation 
we can only compare with valence
quark distributions. But even these will be influenced by the presence
(or absence) of the sea. In fact, the phenomenological values indicated
by the asterisks in Fig.\ref{fig:proton}
differ substantially from the extrapolated
lattice data. One might however expect that the flavour-nonsinglet
combination $\langle x \rangle ^{(u)} - \langle x \rangle ^{(d)}$ is
less sensitive to sea quark effects. Indeed our result of 0.23(3)
($\overline{\mbox{MS}}$ scheme)
obtained by averaging $v_{2,a}$ and $v_{2,b}$ 
compares more favourably with the phenomenological number 0.18.  

On the other hand, the large
discrepancy between the quenched and the phenomenological values of 
$\langle x \rangle ^{(q)}$ should not be too surprising. In the real
proton the contributions of the valence quarks, the sea quarks, and
the gluons must add up to 1. In the quenched approximation, the
missing sea contribution, which is about 0.18, 
must somehow be compensated by the valence
quarks and the gluons. Assuming that the gluon distribution is not too
strongly affected by quenching one would expect the quenched 
$\langle x \rangle ^{(q)}$ to be larger than the phenomenological
result, which is exactly what we find. 

The values obtained for $\langle x^2 \rangle$ and $\langle x^3 \rangle$
($\overline{\mbox{MS}}$ scheme)
differ less drastically from their phenomenological counterparts:
\begin{center}
\begin{tabular}{cr@{.}lr@{.}lr@{.}lr@{.}l}
{} & \multicolumn{4}{c}{lattice} & \multicolumn{4}{c}{experiment} \\
{} & \multicolumn{2}{c}{$u$} & \multicolumn{2}{c}{$d$} & 
     \multicolumn{2}{c}{$u$} & \multicolumn{2}{c}{$d$} \\
 $\langle x \rangle_a $    &  0&369(26) & 0&169(13) & 0&284 & 0&102 \\
 $\langle x \rangle_b $    &  0&440(22) & 0&187(10) & 0&284 & 0&102 \\
 $\langle x^2 \rangle $    &  0&108(16) & 0&036(8)  & 0&083 & 0&025 \\
 $\langle x^3 \rangle $    &  0&020(10) &-0&001(6)  & 0&032 & 0&008 
\end{tabular}
\end{center}

\section{RESULTS FOR POLARISED NUCLEON STRUCTURE FUNCTIONS}

\begin{figure}[t]   
\vspace*{-1.4cm}
\hspace*{-1cm}
\epsfig{file=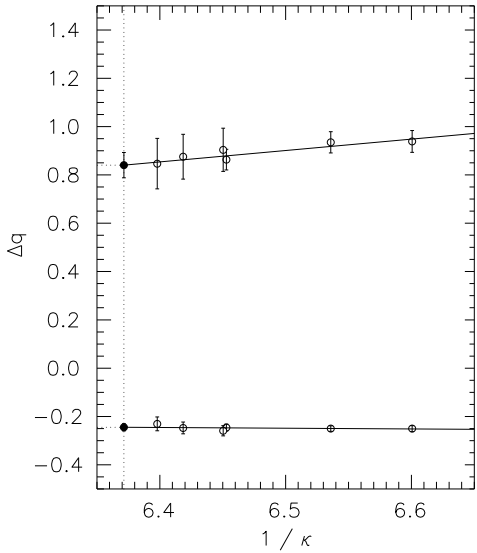,width=9cm}
\vspace*{-4.2cm}
 \caption[dummy]{$\Delta u$ (upper values) and $\Delta d$ (lower values)
         for the proton.} 
\label{fig:deltaq}
\end{figure}

As in the case of the unpolarised structure functions, 
the chiral extrapolation of
the $16^3 \times 32$ data \cite{letter} is confirmed by the results
obtained on the $24^3 \times 32$ lattice at smaller quark masses. 
This is shown in Fig.\ref{fig:deltaq} for the case of $\Delta q$.
In the chiral limit we find 
\begin{equation}
 \Delta u = 0.84(5) \,,\, \Delta d = -0.24(1) \,. 
\end{equation}

These numbers should be compared with the fraction of the proton spin
carried by the valence quarks 
(see, e.g., Ref.\ \cite{cheng}):
\begin{equation}
 \Delta u _v = 0.92 \,,\, \Delta d _v = -0.34 \,. 
\end{equation}
Again one might hope that the difference of $u$ and $d$ contributions
is less sensitive to quenching. One finds $\Delta u - \Delta d = 1.08(6)$ 
to be compared with the axial vector coupling constant $ g_A = 1.26$.

In the higher moments of $g_1$ sea quark effects are also expected to
be suppressed so that the quenched results should be reasonably close
to the experimental numbers. For the proton we find 
\begin{equation}
  \int_0^1 dx \, x^2 g_1 (x,Q^2) = 0.0150(32) \,,
\end{equation}
where $Q^2 = \mu ^2 \approx 5 \mbox{GeV}^2$. The E143 collaboration
obtains in a recent analysis 0.0121(10) \cite{stuart}.

\section{PION STRUCTURE FUNCTIONS}
 
\begin{figure}[t]   
\vspace*{-1.4cm}
\hspace*{-1cm}
\epsfig{file=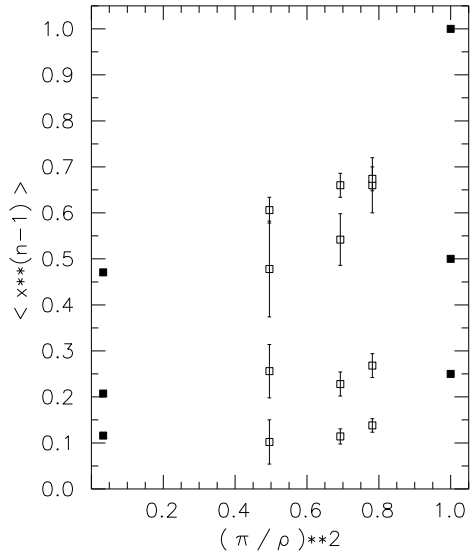,width=9cm}
\vspace*{-4.2cm}
 \caption[dummy]{$v_n = \langle x ^{n-1} \rangle$ for the pion 
       versus $(m_\pi / m_\rho)^2$ for $n=2,3,4$ (from top to bottom).}
\label{fig:pion}
\end{figure}

In Fig.\ref{fig:pion} we summarise our results for the valence quark
distribution in the pion obtained on a $16^3 \times 32$ lattice \cite{best}.
We used the same operators as in the nucleon case. 
Like for $\langle x \rangle ^{(u)}$ in the proton,
$v_{2,b}$ is consistently larger than $v_{2,a}$. 
The three filled squares on the right represent the heavy quark limit,
whereas those on the left are calculated from a phenomenological
valence quark distribution \cite{sutton}. 

\section{FIRST RESULTS FROM AN IMPROVED ACTION}
 
Wilson's fermion action suffers from $O(a)$
lattice artifacts. In order to suppress them at least in on-shell
quantities, Sheikholeslami and Wohlert \cite{wohlert}
added the so-called clover term
\begin{equation}
  \frac{\mbox{i}}{4} c_{SW} a \sum_x \bar{q} (x) \sigma_{\mu \nu}
    F_{\mu \nu}(x) q(x) \,
\end{equation}
to the action, 
where $F_{\mu \nu}$ is the clover-leaf lattice version of the field strength.
For the coefficient $c_{SW}$ one obtains in lowest-order perturbation 
theory $c_{SW} = 1$. A nonperturbative determination is however possible 
as the ALPHA collaboration has shown \cite{alpha}. For 
$\beta = 6.0$ they find the optimal value $c_{SW} = 1.769$
\cite{luscher}, which should reduce the discretisation errors from
$O(a)$ to $O(a^2)$. Since a ``canonical'' value of $c_{SW}$ has not
yet emerged, we have extended our perturbative calculation of
renormalisation constants for the local 2-quark operators to the  
Sheikholeslami-Wohlert action with arbitrary $c_{SW}$ \cite{holger}.

In order to improve matrix elements such as those
considered here one needs in addition improved operators. We have
obtained first results with perturbatively improved operators on a
$16^3 \times 32$ lattice at $\beta = 6.0$ and $c_{SW} = 1.769$
\cite{stephenson}. 
The value of $\Delta u - \Delta d$ in the chiral limit changes 
to 1.22(14) for $c_{SW} = 1.769$ leading to
better agreement with the experimental result 1.26.

\section{CONCLUSIONS}

This talk has described recent attempts to calculate hadron structure
functions on the lattice. 
In the quenched approximation, at least the lower moments can be calculated
with reasonable statistical accuracy. 

Concerning the systematic
uncertainties we can make the following statements:  
\begin{itemize}

\item 
The comparison of results from $16^3 \times 32$ and $24^3 \times 32$
lattices does not reveal large finite size effects. 

\item 
The extrapolation to the chiral limit seems to be smooth. In the range
of quark masses that we studied we did not observe any unexpected mass
dependence with the possible exception of $d_2$.

\item
As a check for cut-off effects we calculated $\langle x \rangle$ from
two different lattice operators. 
The outcome indicates that these effects are not too large. 
First results obtained with a
nonperturbatively improved fermion action look promising. 
\end{itemize}

Still there are further sources of systematic errors due to the
renormalisation constants, the contributions of purely gluonic
operators and fermion-line disconnected parts of 2-quark operators,
etc. A major problem is, of course, the quenched approximation.
However, the overall agreement with the real world is already rather
satisfactory, at least as far as quantities are concerned which are
expected to be less sensitive to quenching.

\section*{ACKNOWLEDGEMENTS}
 
This work was supported in part by the Deutsche
Forschungsgemeinschaft. The numerical calculations were performed on
the Quadrics parallel computers at Bielefeld University and at DESY
(Zeuthen). We wish to thank both institutions for their support.

\end{document}